\documentclass[aps,pre,twocolumn,showpacs,floatfix,superscriptaddress]{revtex4}
\usepackage{graphicx}

\newcommand {\be} {\begin{equation}}
\newcommand {\ee} {\end{equation}}
\newcommand {\Be}{\begin{eqnarray*}}
\newcommand {\Ee} {\end{eqnarray*}}
\newcommand {\bey} {\begin{eqnarray}}
\newcommand {\eey} {\end{eqnarray}}
\newcommand{\bit}{\begin{itemize}}      
\newcommand{\eit}{\end{itemize}}
\newcommand{\bfl}{\begin{flusleft}}
\newcommand{\efl}{\end{flusleft}}
\newcommand{\bfr}{\begin{flushright}}
\newcommand{\bc}{\begin{center}}
\newcommand{\ec}{\end{center}}
\newcommand{\ben}{\begin{enumerate}}    
\newcommand{\een}{\end{enumerate}}

\begin{document} 

\title{Exploring the energy landscape of model proteins: 
a metric criterion for the determination of dynamical connectivity} 

\author{Lorenzo Bongini}
\email{bongini@fi.infn.it}
\affiliation{Dipartimento di Fisica, Universit\'a di Firenze,
via Sansone, 1 - I-50019 Sesto Fiorentino, Italy}
\affiliation{Centro Interdipartimentale per lo Studio delle Dinamiche
Complesse, via Sansone, 1 - I-50019 Sesto Fiorentino, Italy}
\author{Roberto Livi}
\email{livi@fi.infn.it}
\affiliation{Dipartimento di Fisica, Universit\'a di Firenze,
via Sansone, 1 - I-50019 Sesto Fiorentino, Italy}
\affiliation{Centro Interdipartimentale per lo Studio delle Dinamiche
Complesse, via Sansone, 1 - I-50019 Sesto Fiorentino, Italy}
\author{Antonio Politi}
\email{politi@inoa.it}
\affiliation{Istituto dei Sistemi Complessi, CNR, 
L.go E. Fermi 6
I-50125 Firenze, Italy}
\affiliation{Centro Interdipartimentale per lo Studio delle Dinamiche
Complesse, via Sansone, 1 - I-50019 Sesto Fiorentino, Italy}
\author{Alessandro Torcini}
\email{torcini@inoa.it}
\affiliation{Istituto dei Sistemi Complessi, CNR, 
L.go E. Fermi 6 I-50125 Firenze, Italy}
\affiliation{Centro Interdipartimentale per lo Studio delle Dinamiche
Complesse, via Sansone, 1 - I-50019 Sesto Fiorentino, Italy}

\begin{abstract}
A method to reconstruct the energy landscape of small peptides is presented
with reference to a 2d off--lattice model. The starting point is a
statistical analysis of the configurational distances between generic
minima and directly connected pairs (DCP). As the mutual distance of DCP is
typically much smaller than that of generic pairs, a metric criterion can
be established to identify the great majority of DCP. Advantages and limits
of this approach are thoroughly analyzed for three different heteropolymeric
chains. A funnel--like structure of the energy landscape is found in all of
the three cases, but the escape rates clearly reveal that the native
configuration is more easily accessible (and is significantly more
stable) for the sequence that is expected to behave as a real protein.
\end{abstract} 


\maketitle

\section{Introduction}
\label{uno}
Several states of matter are characterized by a rich energy landscape (EL),
which, in turn, hints at peculiar structural and dynamical features.
Supercooled liquids, glasses, atomic clusters and biomolecules \cite{wales} are
typical examples of systems whose complex thermodynamic behavior can be
traced back to the intricate topological properties of the EL. The pioneering
work by Stillinger and Weber on ``inherent'' structures of liquids \cite{still2}
revealed the importance of investigating the stationary points of the EL
for characterizing their dynamical and thermodynamic properties. Similar
approaches have been proposed and successfully applied to the identification of
the structural--arrest temperature in glasses \cite{sastry} and supercooled
liquids \cite{angelani}.

More recently, this kind of analysis has been extended to the study of protein
models \cite{krivov,evans,baumketner}. They suggest that also the folding process
of a protein towards its native configuration (NC) depends on the structure of
its EL. This has been found to possess a funnel--like shape: the NC is located
inside the so--called native valley (NV) at the bottom of the funnel \cite{funnel}.

Below the folding temperature, the evolution from a coil state to the NC is
determined by the propensity of the protein to enter the relatively small
fraction of states composing the NV without having to visit the
entire phase-space. In a statistical sense, the folding process can be viewed
as a weighted sampling mechanism which favours specific intermediate
configurations. They correspond to assembling the building blocks which
eventually constitute the NC. Well above the folding temperature, no marked
difference exists among the various states and the protein spends most of the
time jumping between different random coil configurations. 

In the absence of external forces, only thermal fluctuations can drive the
protein dynamics through different regions of the EL. In particular, below the
folding temperature, the protein is expected to evolve mainly inside the NV.  
Nonetheless, large deviations from the NC cannot be avoided, but they
are both rare and very short lived. This scenario was confirmed by simulations
performed in a 2d off--lattice model\cite{tlp}. A more detailed analysis of the
model \cite{tap} revealed that the protein dynamics can viewed as a sequence of
jumps between pairs of minima separated by one saddle, that we call directly
connected pairs (DCP). Each jump is a thermally activated process: the protein
performs random oscillations in the basin of a local minimum until a
sufficiently large thermal fluctuation allows it to overtake the energy barrier
separating the minimum from a neighbouring one. In a high--dimensional space,
the transition rate can be computed as the product of the Arrhenius factor
times an entropic weight which depends on the DCP and on the saddle curvatures
(see Section \ref{quattro}). In other words, the relevant information about the
protein dynamics can be obtained from the knowledge of the DCP and of the
corresponding saddles. However, the reconstruction of the EL is a very
difficult task to be accomplished in practice. Indeed, the identification of
DCP by a systematic exploration of the entire set of the $N$ identified minima
requires to exploring $N^2$ pairs, which is already on the order of $10^9$ for
the partial database generated (see the Appendix for a description of the
algorithm) in the relatively small polypeptidic chains investigated in this
paper (see section \ref{due}). It is therefore, crucial to develop effective
strategies for identifying DCP within the set of all, a priori possible,
candidates. This is the main issue addressed in this paper.

It is reasonable to conjecture that the distance separating DCP is typically
smaller than that between generic pairs of minima. It is therefore tempting to
restrict the analysis to those pairs whose mutual distance is smaller than some
prescribed threshold. However, whether this approach effectively works may
depend on several factors one of which is the adopted definition of distance.
For this reason, in section \ref{tre} we introduce and compare different 
conformational distances. It turns out that the bond--angle distance, defined
by the absolute--value norm is the one that makes the implementation of such a
criterion most accurate in the identification of DCP. The price one has to
pay is that, unavoidably, all DCP characterized by a distance larger than the
given threshold are missed. The detailed analysis of the EL dscribed in Section
\ref{quattro} indicates that the average distance between DCP is larger for
minima that are closer to the NC. Accordingly, the computational advantage
guaranteed by the choice of a relatively small threshold could be frustrated by
the loss of important connections located in the NV. For this reason,
we argue that the distance criterion has to be complemented by a systematic
search of all DCP involving minima in the NV, a much more accessible task,
given the limited number of such minima (for an operative definition of the NV
see section \ref{quattro}).

\section{The model}
\label{due}
The model studied in this paper is a modified version of the 2d off-lattice
model introduced by Stillinger {\it et al.} \cite{still} and already
investigated in Ref.~\cite{tlp}. It consists of a chain of $L$ point-like
monomers mimicking the residues of a polypeptidic chain. For the sake of
simplicity, only two types of residues are considered: hydrophobic (H) and
polar (P) ones. Any chain is unambiguously identified by a sequence of binary
variables $\{ \xi_i \}$ ($i=1, \dots, L$), where $\xi_i = \pm 1$ corresponds to
H and P residues, respectively.
The intramolecular potential is composed of three terms: a stiff
nearest-neighbour harmonic potential, $V_1$, intended to maintain the bond
distance almost constant, a three-body interaction $V_2$, which accounts for the
bending energy and a long--range Lennard-Jones interaction, $V_3$,
acting on all pairs $i$, $j$ such that $|i-j| >1$
\begin{eqnarray} 
V_1 (r_{i,i+1}) &=& \alpha (r_{i,{i+1}}-r_0)^2, \nonumber\\
V_2(\theta_i) &=& \frac{1 - \cos\theta_i}{16},\nonumber\\
V_3(r_{i,j}) &=& \frac{1}{r_{i,j}^{12}} - \frac{c_{i,j}}{r_{i,j}^6}
\label{v1}
\end{eqnarray}
Here, $r_{i,j}$ is the distance between the $i$-th and the $j$-th monomer 
and $\theta_i$ is the bond angle at the $i$-th monomer. The parameters
$\alpha =20$ and $r_0 =1$ (both expressed in adimensional arbitrary units) fix
the strength of the harmonic force and the equilibrium distance between
subsequent monomers (which, in real proteins, is on the order of a few \AA).
The value of $\alpha$ is chosen to ensure a value for $V_1$
much larger than the other terms of potential (\ref{v1}) in order 
to reproduce the stiffness of the
protein backbone. $V_3$ is the only term of the potential energy which depends
on the nature of the monomers: the parameters
$c_{i,j} = \frac{1}{8} (1+\xi_i + \xi_j +5 \xi_i \xi_j)$ are chosen in such a
way that the interaction is attractive if both residues are either hydrophobic
or polar ($c_{i,j} = 1$ and $1/2$, respectively), while it is repulsive if the
residues belong to different species ($c_{ij} = -1/2$).
\newline
Accordingly, the Hamiltonian of the system reads
\begin{eqnarray}
H = \sum_{i=1}^L \frac{p_{x,i}^2+p_{y,i}^2}{2} +\sum_{i=1}^{L-1} V_1(r_{i,i+1}) +
\nonumber\\
+\sum_{i=2}^{L-1} V_2(\theta_i)+ \sum_{i=1}^{L-2} \sum_{j=i+2}^{L}  V_3(r_{ij},\xi_i,\xi_j)
\label{hamil}
\end{eqnarray}
where, for the sake of simplicity, all monomers are assumed to have the same
unitary mass and the momenta are defined as
$(p_{x,i},p_{y,i}) \equiv ({\dot x}_i,{\dot y}_i)$.
Despite its simplicity, this toy--model of a heteropolymer does reproduce the
expected properties of polypeptidic chains and is thus very useful for
testing dynamical and statistical indicators. For instance, accurate
Monte-Carlo simulations, performed by employing innovative
schemes~\cite{marinari}, have revealed that, in analogy with real proteins,
only a few sequences systematically fold onto the same native structure: this
is why they have been named ``good folders" \cite{irback1,irback2}. Such studies
have been confirmed and complemented by direct molecular dynamics
simulations~\cite{tlp}.

In this paper we limit ourselves to investigating the three following sequences
of twenty monomers,

\begin{itemize}

\item{[S0]} a homopolymer composed of 20 H residues;

\item{[S1]=[HHHP HHHP HHHP PHHP PHHH]} a sequence that has been
identified as a good folder in \cite{irback2};

\item{[S2]=[PPPH HPHH HHHH HHHP HHPH]} a randomly generated sequence,
that has been identified as a bad folder in \cite{tlp}.

\end{itemize}

These sequences have been chosen because they represent the three classes 
of different folding behaviors observed in this model. The main thermodynamic
features of all of them can be summarized with reference to three different 
transition temperatures \cite{tap}. Decreasing the temperature from high values,
one first encounters the temperature $T_\theta$ below which the sequence is
typically found in a collapsed configuration, rather than in a random-coil one
\cite{degennes}. Then, one finds the so--called folding temperature $T_f$ below
which the heteropolymer stays predominantly in the native valley.  Finally, at
even lower temperatures one finds $T_g$: this is the glass-transition
temperature, below which a structural arrest of the system occurs.

In the following we aim at a deeper understanding of the folding process by
investigating directly the EL structure of all these sequences.

\section{Analysis of the energy landscape}
\label{tre}

As pointed out in the introduction, the main problem for reconstructing
the EL amounts to finding DCP and the corresponding saddles of the potential
energy $V = V_1+V_2+V_3$ (see Eq.(\ref{hamil})). An exhaustive search of DCP
among all pairs of minima rapidly becomes unfeasible with increasing the chain
length $L$, due to the exponential increase of the number of minima with $L$
itself \cite{wales}. Since the total number of DCP is a rather small fraction of all
possible pairs (see, e.g., the Table in the Appendix), it would be very
helpful finding an effective criterion to restrict the search of potentially
directly connected minima. A priori, the {\it distance} seems to be the right
indicator to discriminate between connected and not-contiguous pairs of minima.
In this section we investigate several definitions of distance with the goal
of identifying the most appropriate one to identify DCP.

As a first candidate, we introduce the generalized angular distance
$\delta_{{\theta}}^{(q)}({\cal C}_1,{\cal C}_2)$ between configurations
${\cal C}_1$ and ${\cal C}_2$,
\begin{equation}
\delta_{{\theta}}^{(q)}({\cal C}_1,{\cal C}_2)=
\left(\frac{1}{L-2}
\sum_{i=2}^{L-1}|\theta(i;{\cal C}_1)-\theta(i;{\cal C}_2)|^q\right)^{1 / q} ,
\label{distang}
\end{equation}
where $\theta(i,{\cal C})$ is the $i$-th bond angle of the configuration
$\cal C$. Notice that this angular distance is much sensitive to local
flucutuations along the chain. A generalized distance which depends more on the
global than on the local structure of a configurations is
\begin{equation}
\delta_{r}^{(q)}({\cal C}_1,{\cal C}_2)=
\left(\frac{2}{L(L-1)}\sum_{i>j+1}^L|r(i,j;{\cal C}_1)-r(i,j;{\cal C}_2)|^q\right)^{1/ q} ,
\label{distglobal}
\end{equation}
where $r(i,j;{\cal C})$ is the intra--bead distance between $i$th and $j$th
monomer of the configuration $\cal C$. This distance is related to the
$\chi$-indicator,
previously employed in the analysis of the folding dynamics in on-- and
off--lattice models of heteropolymers in 2d and 3d \cite{veit,tlp}. For $q=1$, 2,
and $+\infty$, both definitions of generalized distances reduce to the standard
absolute--value, Euclidean and maximum norms, respectively. Such distances have
been computed for all the pairs of minima in the databases of the sequences S0,
S1 and S2. The algorithm used to generate the databases is described in the
Appendix. We want to point out that any numerical procedure, including ours,
cannot guarantee the identification of all minima and saddles in the EL.
Nonetheless, we have independently verified that the algorithm allows
obtaining at least a very accurate description of the native valley.

\begin{figure}[h]
\includegraphics[clip,width=6.5cm]{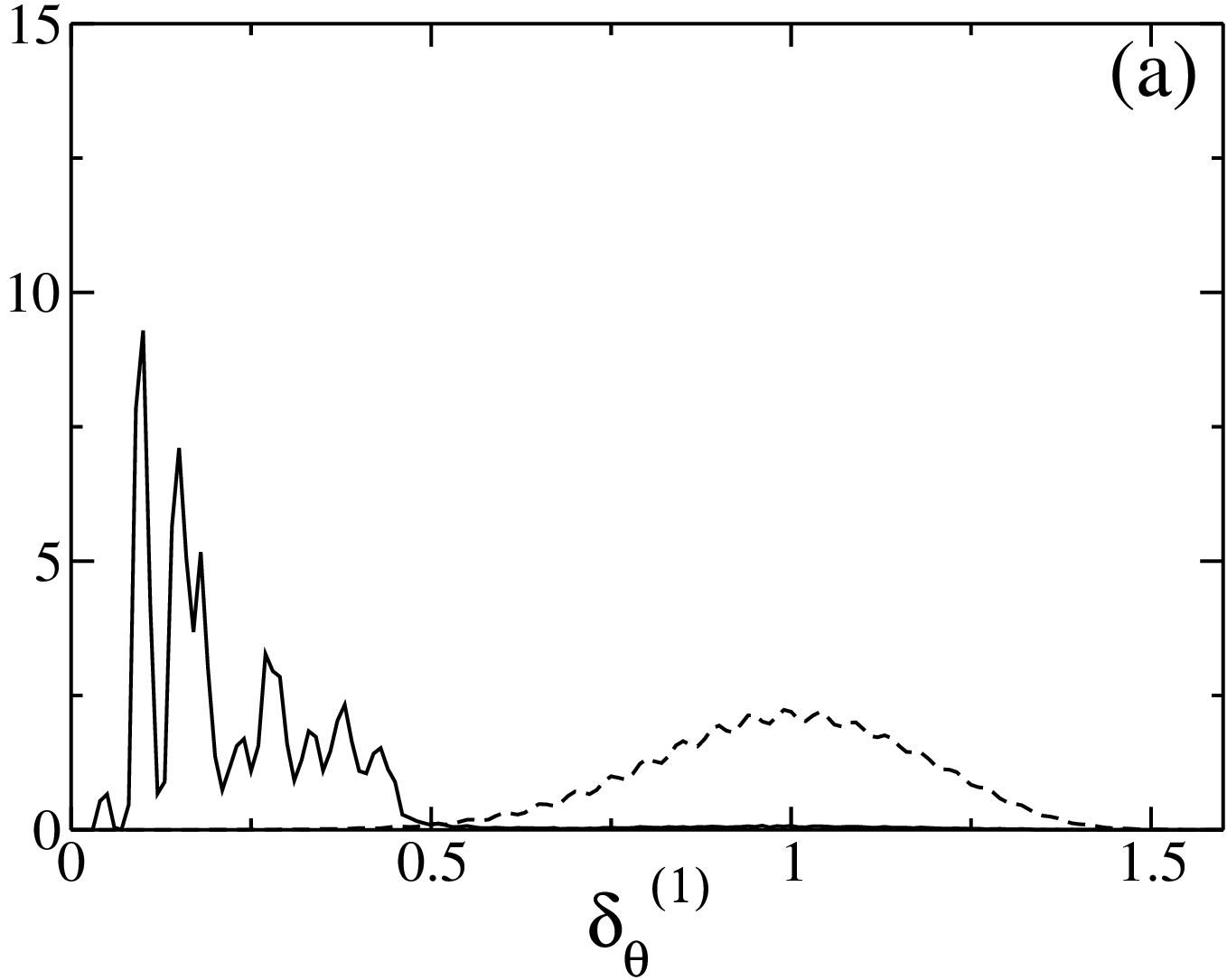}
\includegraphics[clip,width=6.5cm]{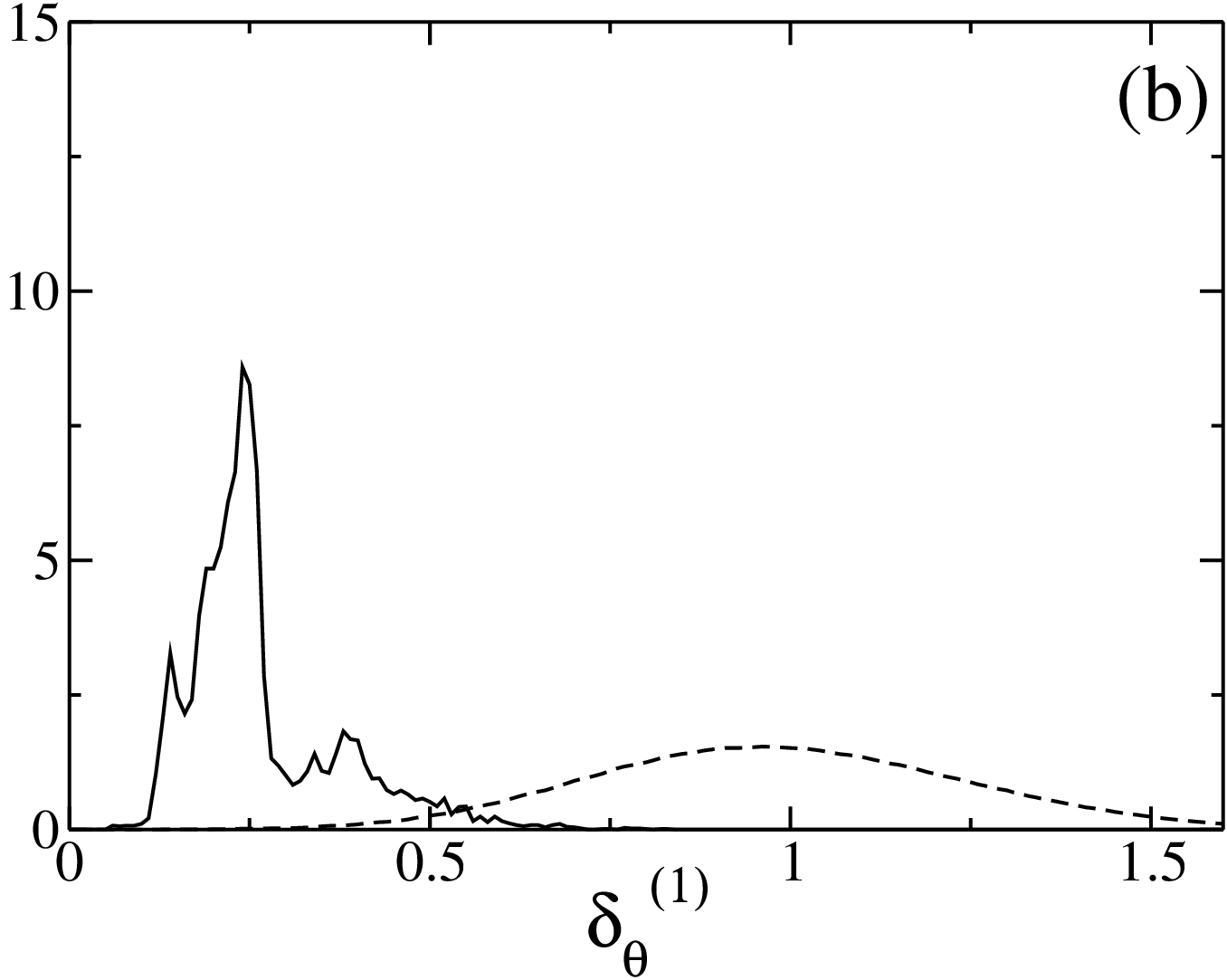}
\includegraphics[clip,width=6.5cm]{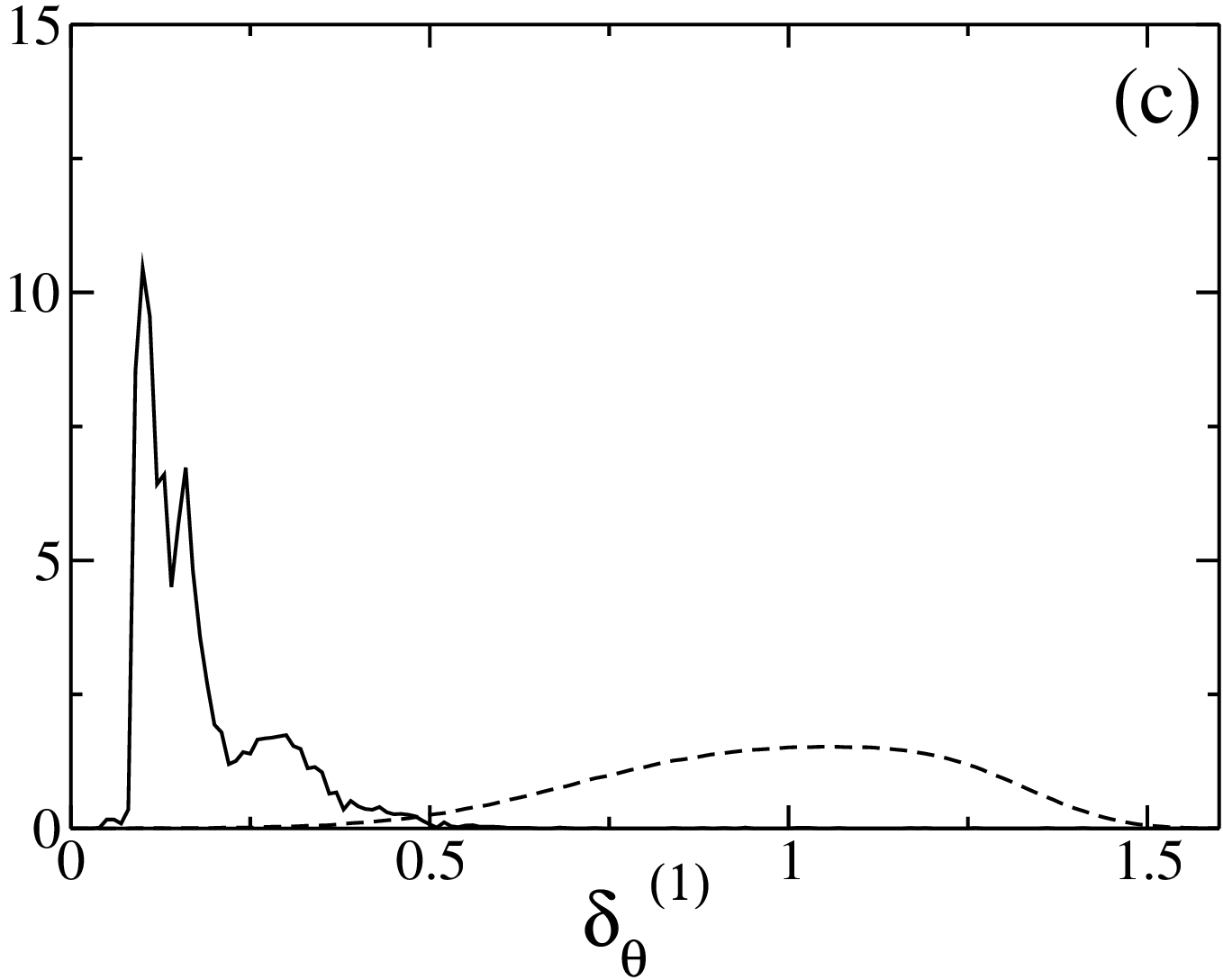}
\caption{Probability density of angular distances $\delta_\Theta^{(1)}$
for all the pairs of minima, $P_a$ (dashed line), and for DCP, $P_c$
(solid line), for the sequences S0 (a) and S1 (b) and S2 (c).}
\label{fig:distrib}
\end{figure}

Then, we have computed the probability densities of the generalized distances
between generic ($P(\delta)$) and directly connected ($P_c(\delta)$) pairs of
minima. In all cases, $P$ has a bell shape with a maximum close to 1, while
$P_c$ is sharply peaked at much smaller values (see, e.g., Fig.~\ref{fig:distrib},
where $P(\delta_\theta^{(1)})$ and $P_c(\delta_\theta^{(1)})$ are plotted 
for the sequences S0, S1 and S2). This confirms the naive idea that DCP are
typically much closer than randomly chosen pairs of minima. 

\begin{figure}[ht!]
\includegraphics[clip,width=6.5cm]{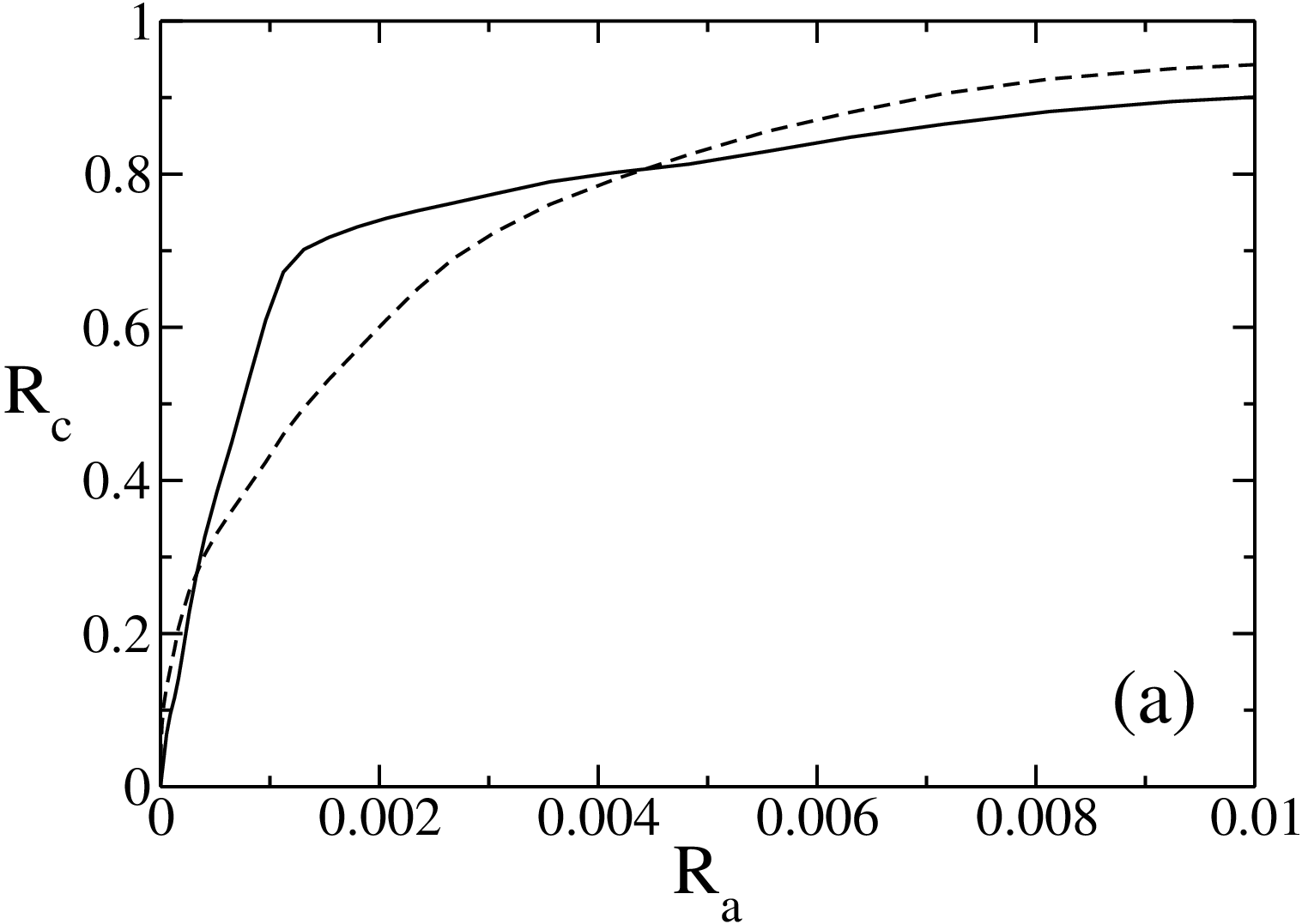}
\includegraphics[clip,width=6.5cm]{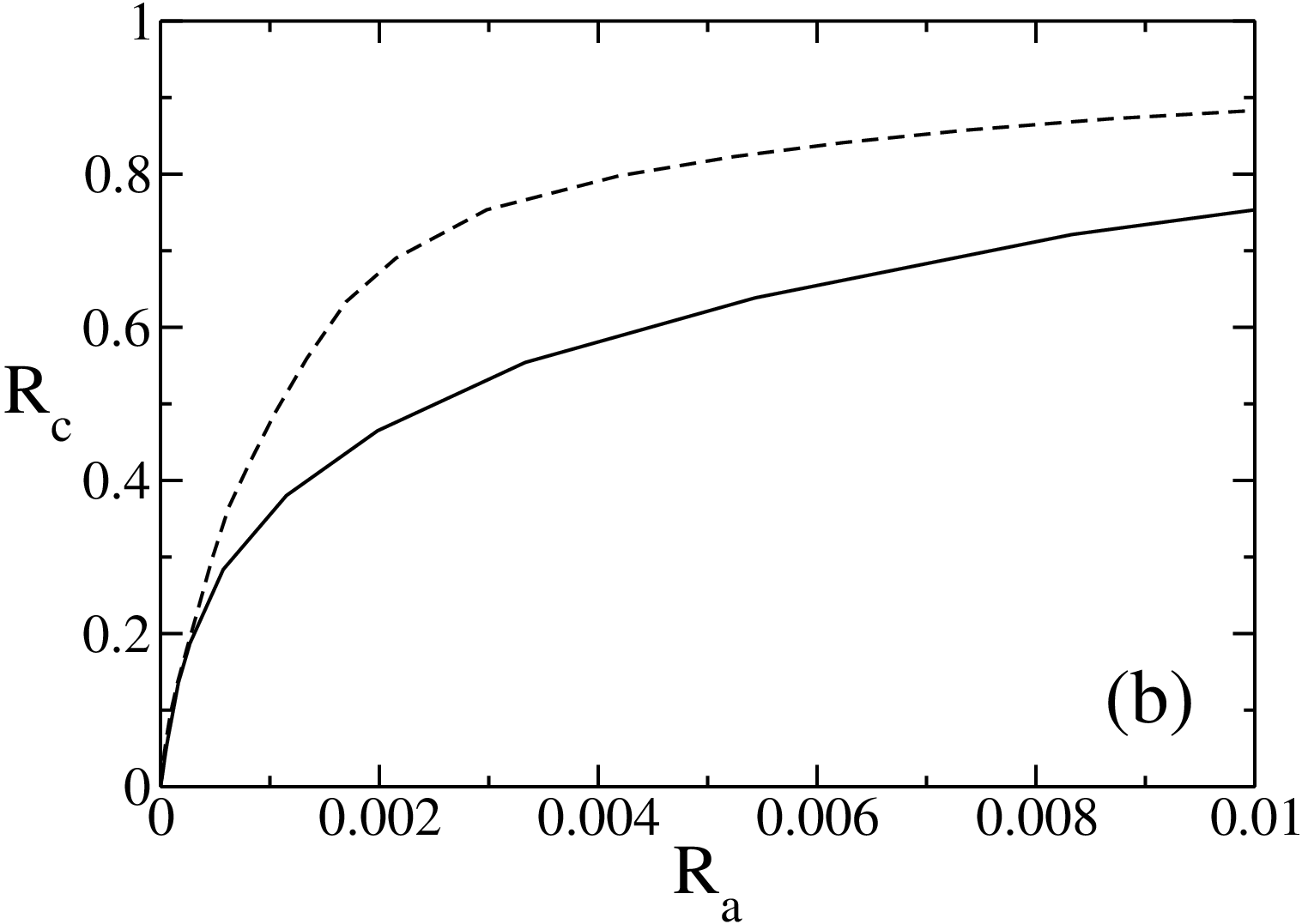}
\caption{Integrated probability density $R_c$ versus $R_a$ for the angular
$\delta_\Theta$ (solid lines) and global $\delta_r$ (dashed lines) distances.
The data refer to the the sequence S1 and to the 
absolute value (a) and maximum (b) norm.}
\label{fig:int_distrib}
\end{figure}

\begin{figure}[ht!]
\includegraphics[clip,width=6.5cm]{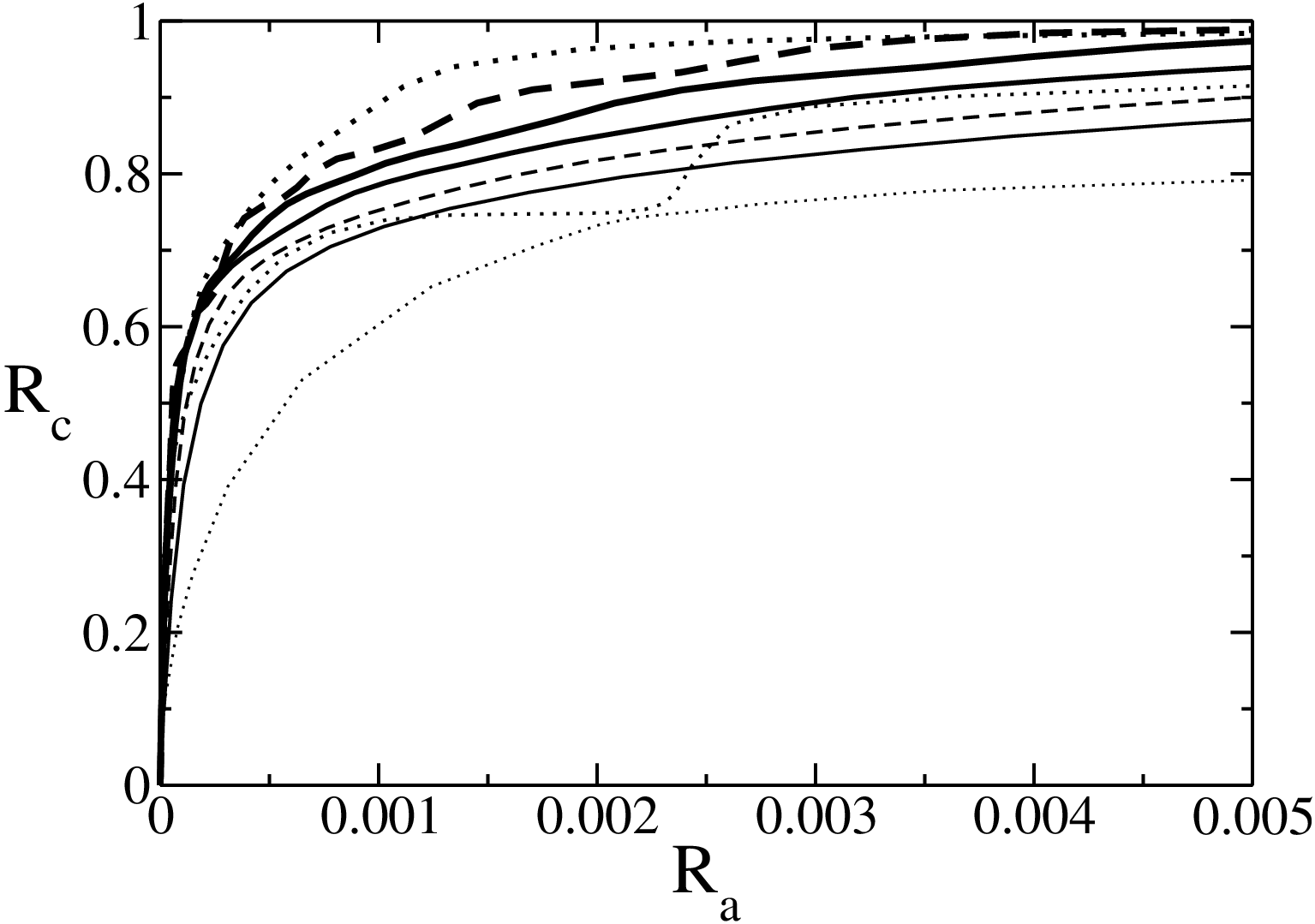}
\caption{
Integrated probability density $R_c$ versus $R_a$ for the angular
distance for various norms for the sequence S0. Contiuous lines correspond to
$q<1$ ($q=0.01$, $q=0.2$, $q=0.5$, and $q=0.75$, going from the thinnest to the
tickest line, respectively). The dashed line corresponds to $q=1$ (norm of
the absolute value). The dotted lines correspond $q>1$ ($q=2$, $q=5$, and 
$q=\infty$, going from the thickest to the thinnest line, respectively).
}
\label{fig:dang00}
\end{figure}

\begin{figure}[ht!]
\includegraphics[clip,width=6.5cm]{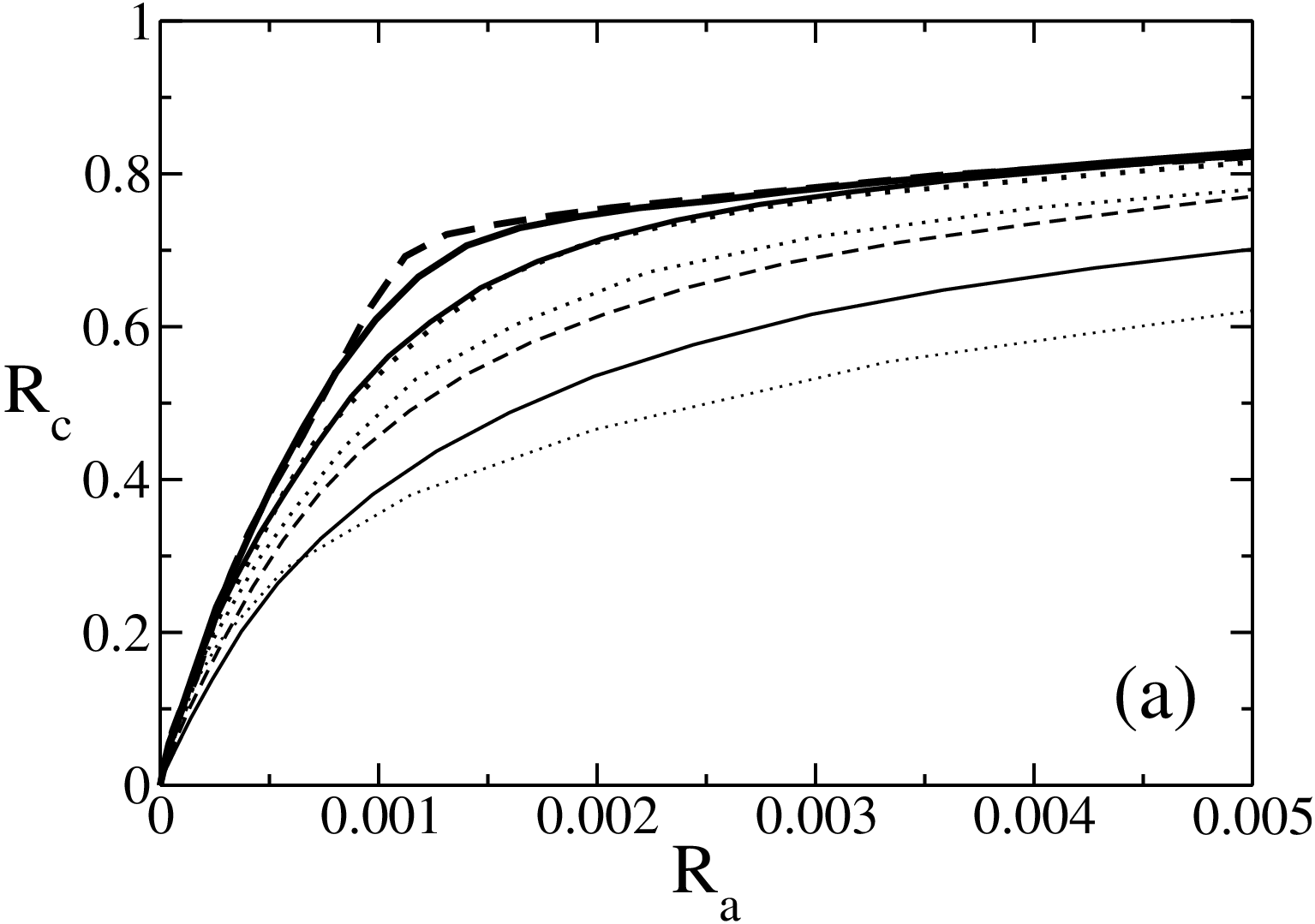}
\includegraphics[clip,width=6.5cm]{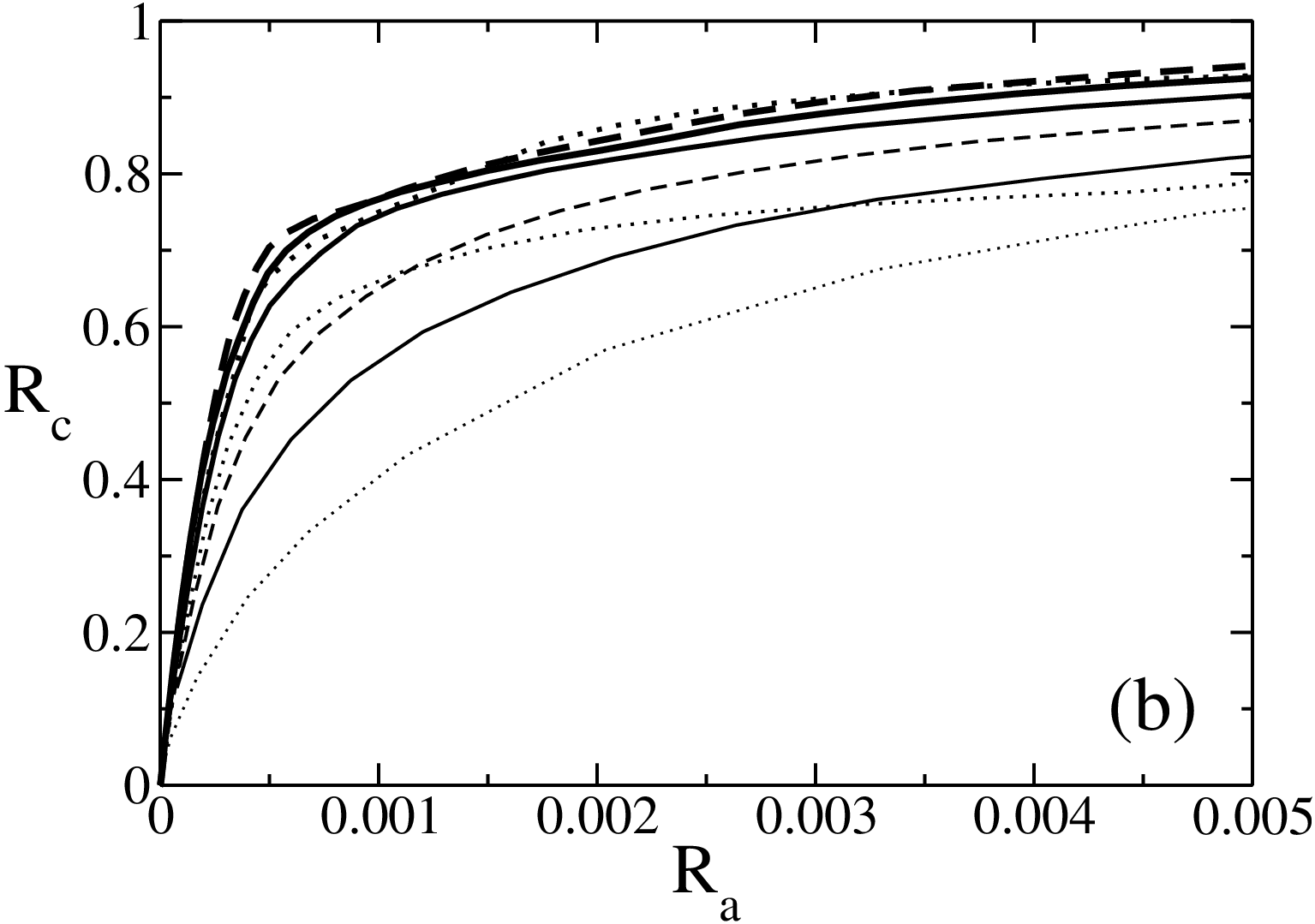}
\caption{
Integrated probability density $R_c$ versus $R_a$ for the angular
distance for various norms for the sequences S1 (a) and S2 (b).
The notations are the same as in Fig.\ref{fig:dang00}}
\label{fig:dang_etero}
\end{figure}

A qualitatively similar difference between $P$ and $P_c$ is observed also
for different choices of $q$ as well as for the global distance
$\delta_r^{(q)}$. In order to identify the most appropriate value of $q$,
it is convenient to introduce the integrated fraction $R(\delta)$ of pairs of
minima whose distance is smaller than $\delta$
\begin{equation}
R(\delta) = \int_0^{\delta} P(x) {\rm d} x 
\label{rate_c}
\end{equation}
and, equivalently,
\begin{equation}
R_c(\delta) = \int_0^{\delta} P_c (x) {\rm d} x \, ,
\label{rate_a}
\end{equation}
relative to DCP. Upon considering $\delta$ as a dummy variable, it is possible
to plot $R_c$ versus $R$. A fast convergence of $R_c$ to 1 means that almost all
DCP can be already identified by limiting the search to relatively close pairs
of minima. The data plotted in Fig.\ref{fig:int_distrib}a for S1  and $q=1$
indeed reveal such a fast growth of $R_c$, that almost $90 \%$
of the DCP can be obtained by investigating only $1\%$  of all pairs both
by considering the angular $\delta_\theta^{(q)}$ and the global $\delta_r^{(q)}$
distance. The curves in Fig.\ref{fig:int_distrib}b show that significantly
worse performances are obtained when the maximum norm ($q=\infty$) is used
to classify the pairs of minima. In order to clarify the role of the parameter
$q$, in Figs.\ref{fig:dang00},\ref{fig:dang_etero}, we have plotted $R_c$
versus $R$ for different values of this parameter. There we see that,
upon decreasing $q$ from $\infty$ down to $1$, $R_c$ exhibits an increasingly
fast saturation, while the opposite is observed when $q$ is further decreased
below $1$. The bad performance observed at high $q$-values has a quite intuitive
explanation: in that limit, the norm reduces to the maximum norm and the slow
growth of $R_c$ tells us that distances between DCP are not uniformly small
along all directions: DCP may significantly differ along specific directions in
spite of being ``on the average" much closer than generic pairs of minima. The
relatively bad performance observed for $q \to 0$ has a complementary
explanation. In that limit, the average distance is strongly biased by small
differences $\delta \theta$ or $\delta r$, whose occasional occurrence may
induce to classify as ``close'', configurations that are significantly 
different instead. In all configurations we have investigated, it turns out
that $q \approx 1$ is the best compromise between the above two effects.
Having established that $q=1$ is the best choice, from now on, we limit
ourselves to considering that value and drop the superscript $(1)$ in the
definition of the distance.

\begin{figure}[ht!]
\includegraphics[clip,width=6.5cm]{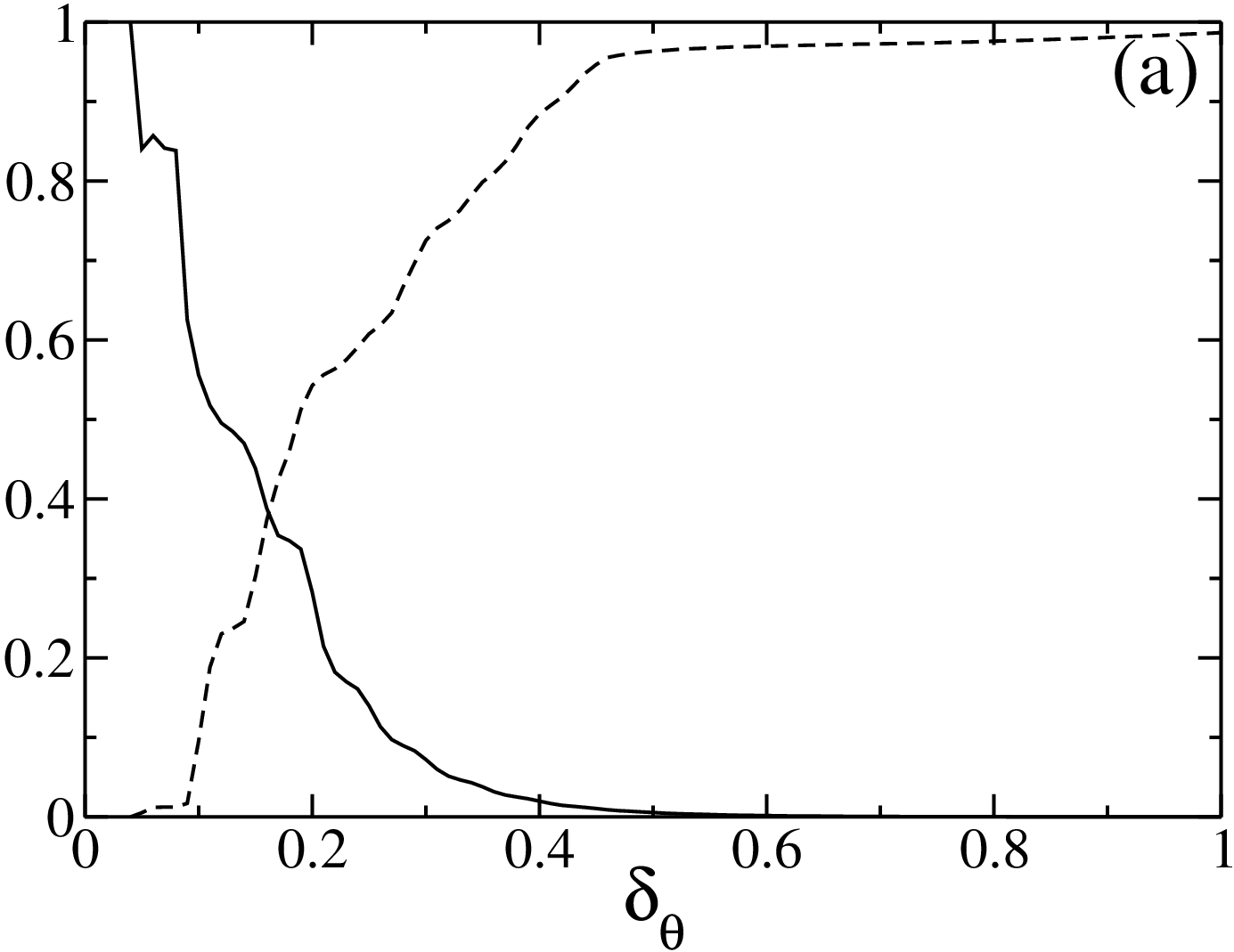}
\includegraphics[clip,width=6.5cm]{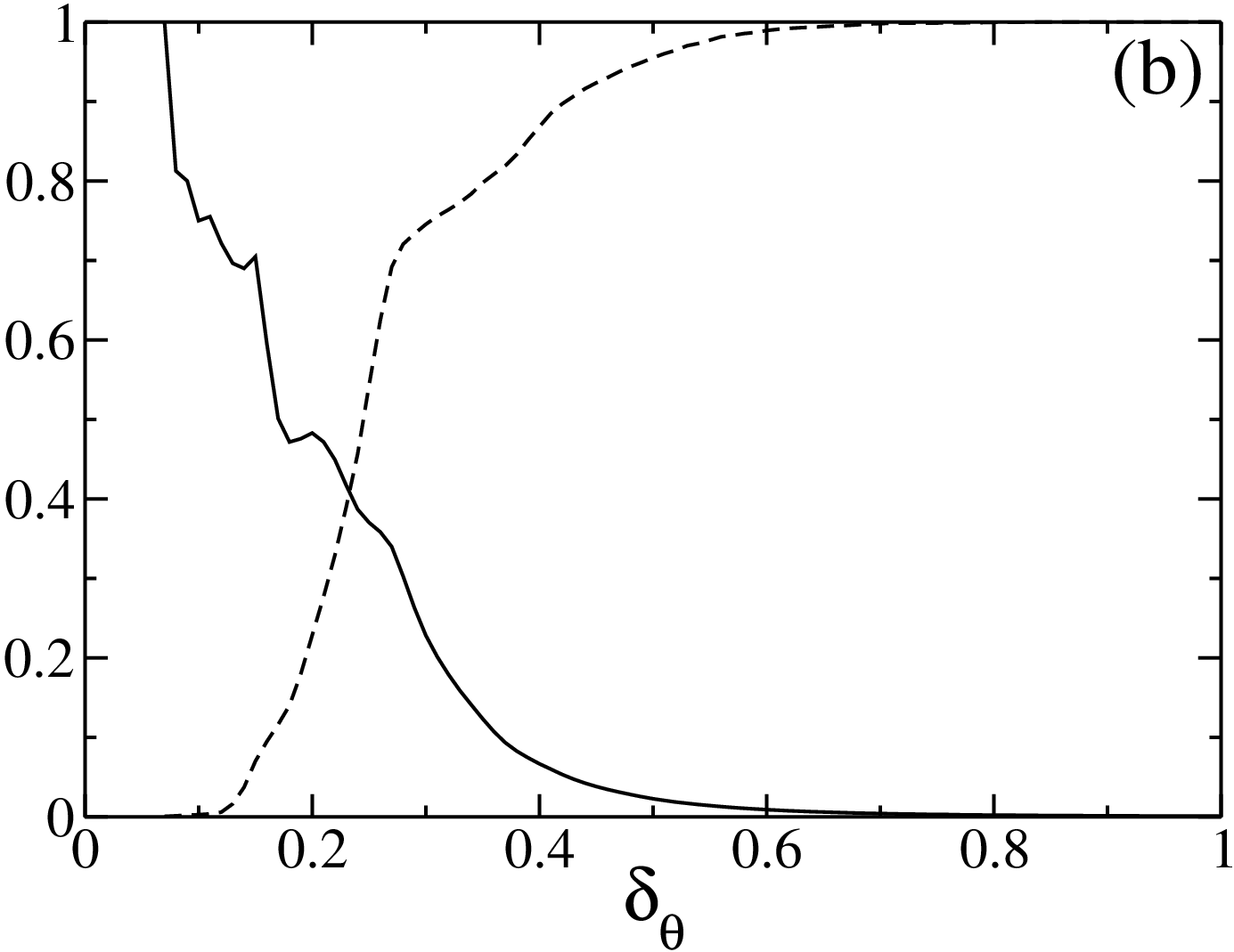}
\includegraphics[clip,width=6.5cm]{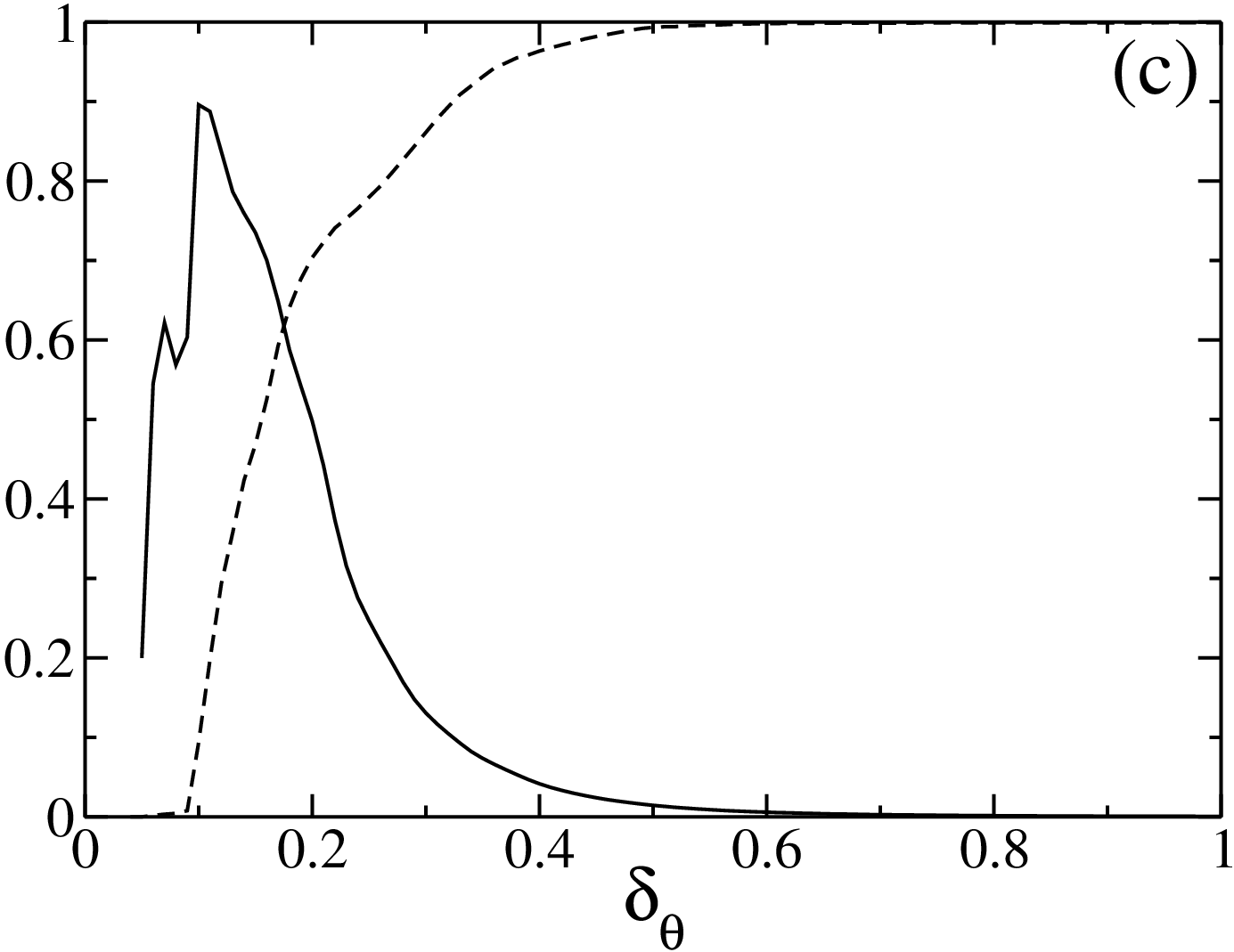}
\caption{
The ratio $\rho$ between DCP and all pairs up to distance 
$\delta_\Theta$ (continuous line) compared with
$R_c(\delta_\Theta)$ (dashed line) 
for the three analyzed sequences, S0 (a), S1 (b) and S2 (c). 
}
\label{fig:rate}
\end{figure}

In practice, since it is eventually necessary to identify a threshold
distance $\delta_\theta^\star$, it is convenient to look also at the
dependence of $R_c$ on $\delta_\theta$ and to introduce
$$
\rho(\delta_\theta) = \frac{R_c(\delta_\theta)}{R(\delta_\theta)}
$$
From Fig.~\ref{fig:rate}, we see that $\delta_\Theta^\star = 0.5$ is a good
choice, since $\rho(0.5) \sim {\cal O}(10^{-2})$, while $R_c(0.5) \sim 99\%$.
Even reducing  the threshold value to $\delta_\theta^\star = 0.2$
a large fraction of DCP are still recovered ($R_c(0.2) \sim 90\%$). 

These results indicate that if one restricts the search of DCP to the set
of minima whose distance is smaller than a prescribed threshold
$\delta_\theta^\star$, one can reduce significantly the most time-consuming part
of the systematic search algorithm, which amounts to testing whether a generic
pair of minima is separated by a single saddle.

The drawback is that for any choice of $\delta_\theta^\star$, those DCP whose
mutual distance is anomalously large are going to be missed. In the next
Section we analyse the EL with the main goal of concluding whether such a
component is of some relevance in the overall reconstruction of the EL.

\section{A closer inspection of the Energy Landscape}
\label{quattro}

A faithful reconstruction of the EL requires a sufficiently large database of
stationary points, i.e. minima and saddles. The procedure described in the
Appendix is quite reliable in this respect, but it is also computationally very
time-consuming as already stated. In the previous section we have seen that a
large fraction of DCP can be obtained by adopting a suitable metric criterion.
However, it is not a priori clear whether the long-distance tail of $P_c$ 
is qualitatively irrelevant too.

In order to shed some light on this question we have divided the minima
into ``shells": the $n$--th shell is defined as the collection of all minima
which are separated from the NC by at least $n$ saddles. The identification of
the minima belonging to each shell can be achieved recursively:
\begin{itemize}
\item the 0-th shell coincides with NC;
\item a minimum ${\cal C}_1$, directly connected to a minimimum ${\cal C}_2$ of the
$i$-th shell, is identified as part of the $i+1$-st shell if ${\cal C}_2$ does not
belong to a shell of order $j \le i$.
\end{itemize}

\begin{figure}[h]
\includegraphics[clip,width=6.5cm]{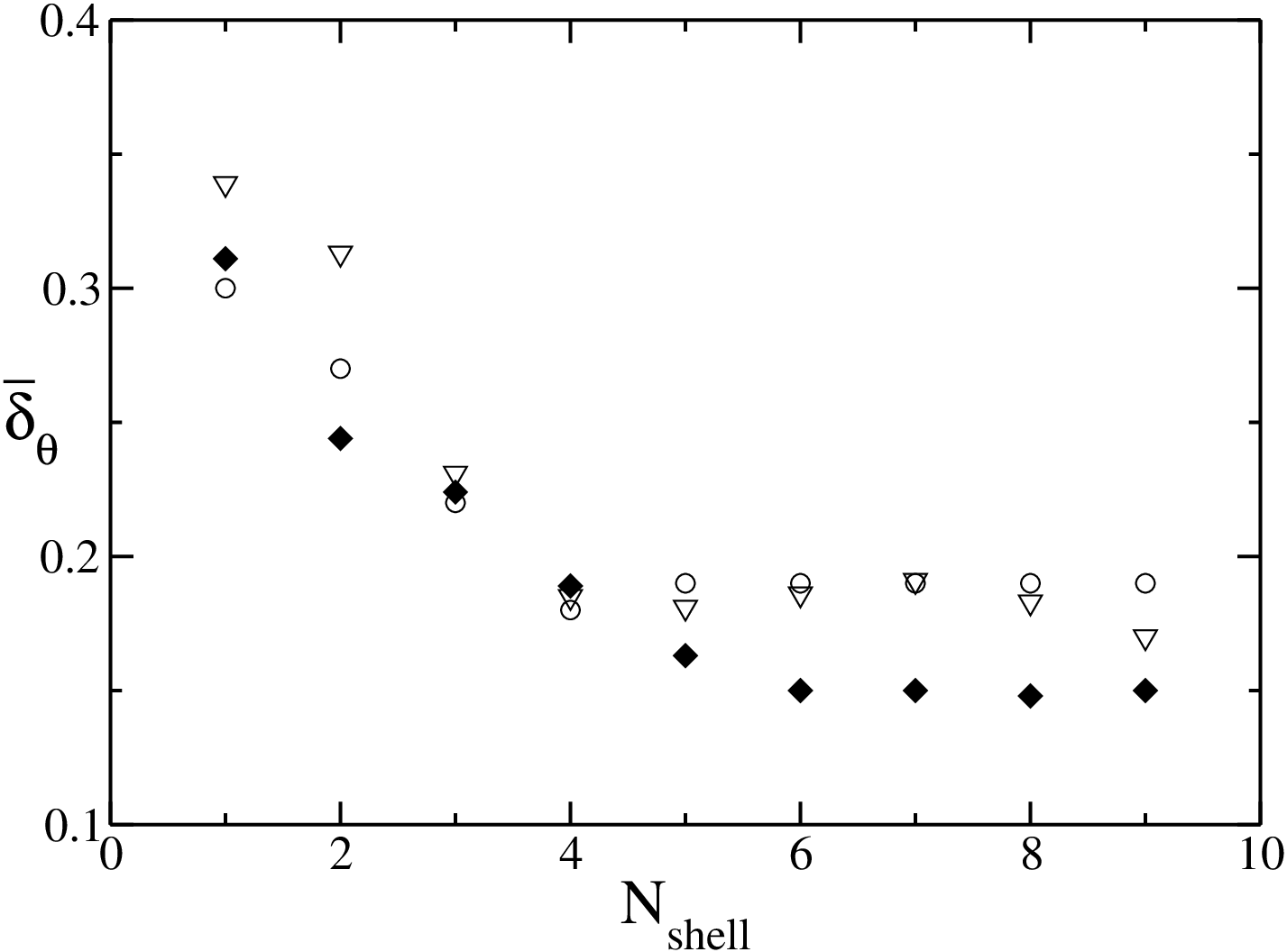}
\caption{Average connection length between the minima of the $i$-th and 
$i-1$-th shell for the three sequences S0 (triangles), S1 (circles) and S2
(filled diamonds). 
}
\label{fig:dist_shell}
\end{figure}

\begin{figure}[h]
\includegraphics[clip,width=6.5cm]{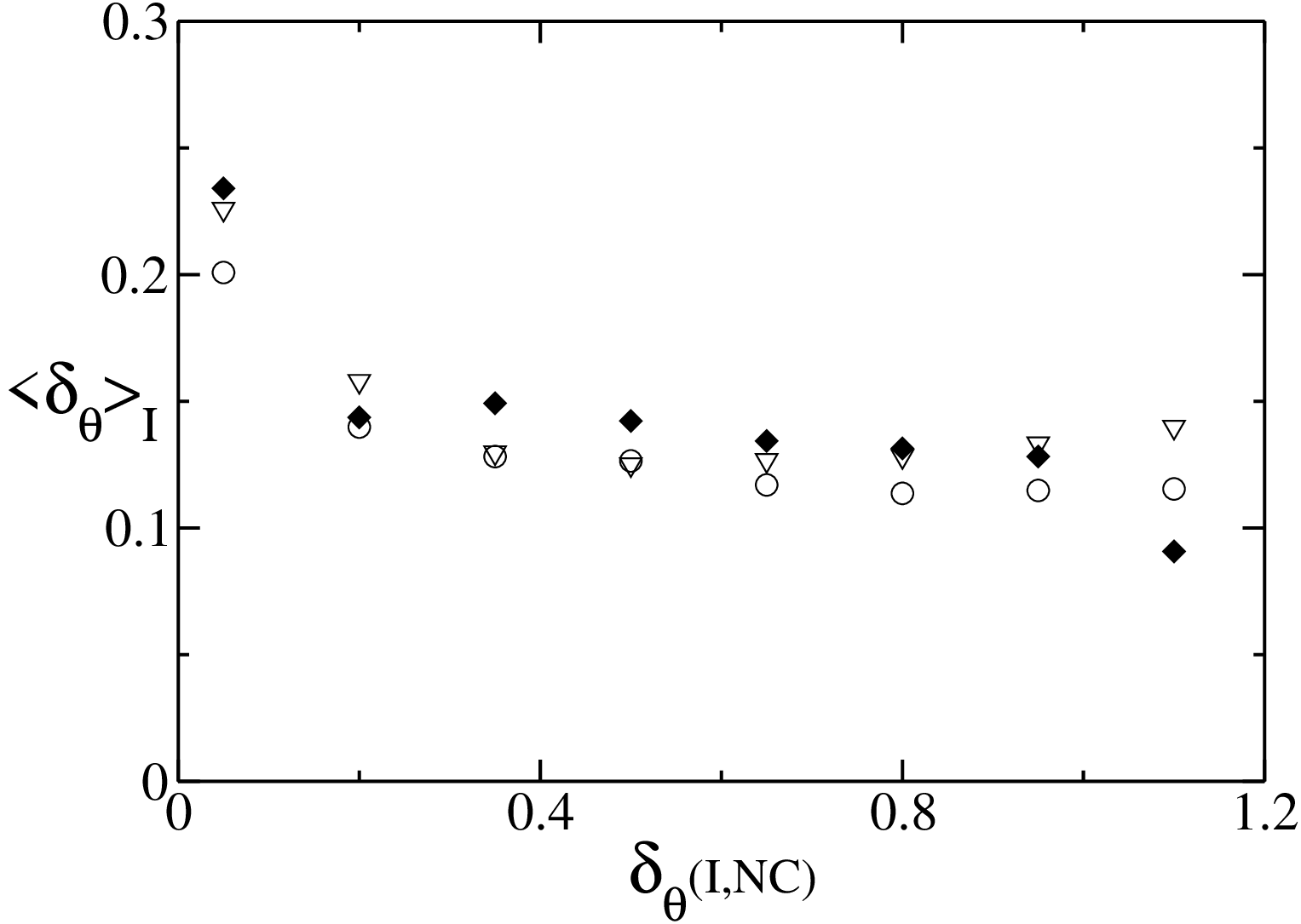}
\caption{
The average distance of minima from their connected set versus
their distance from the NC
for the sequence S0 (triangles), S1 (circles) and S2 (filled diamonds).
The data are averaged also over intervals of length $I=0.01$ along the
horizontal axis.}
\label{fig:dist_nat}
\end{figure}

In Fig.~\ref{fig:dist_shell}, we report the average value $\bar\delta_\theta$ 
of the distance between DCP belonging to consecutive shells for all of the
three sequences. This figure indicates that the interminimum distance grows
in the vicinity of the NC. The average distance $\langle\delta_\theta\rangle$ between DCP lying inside the same shell exhibits the same behaviour. 
Finally, this rarefied density of minima in
the vicinity of the NC is confirmed also by plotting the mutual distance versus
the actual distance from the NC. In this case, in order to smoothen the wild
pair-to-pair fluctuations, a coarse-graining has been performed by averaging
over bins of widht 0.01 along the $\delta_\theta$ axis (see
Fig.~\ref{fig:dist_nat}).

Altogether, the significative increase of the average distance close to the NV
indicates that some of the DCP that are missed by the metric criterion
discussed in previous section lie in the most relevant region for the
characterization of the folding/unfolding processes. However, considering the
limited number of minima in the NV, such a difficulty can be easily overcome by
complementing the overall application of the metric criterion with an extensive
comparison of such minima with all configurations in the database: the
additional cost in terms of the computing time is indeed a minor one.

The increased distance between neighbouring minima hints at possibly deeper
valleys and is, in turn, suggestive of the presence of a funnel in the EL, a
structure that is typically expected to appear in true proteins \cite{funnel}. 
However, we
find this scenario in all of the three sequences analyzed in this paper,
including the homopolymer S0 which cannot be certainly considered a reasonable
model for a protein. In order to further clarify this point, we have computed
the escape rates from the single valleys. Given any two directly connected
minima ${\cal C}_1$ and ${\cal C}_2$ characterized by the potential energies
$V_1$ and $V_2$ ($V_1 \leq  V_2$), the escape rate from the minimum
${\cal C} = \{ {\cal C}_1, {\cal C}_2\}$ through the saddle ${\cal C}_s$ (with
potential energy $V_s$) is given by
\begin{equation}
\Gamma_{\cal C} = \frac{\omega_\parallel}{\pi \gamma}
\frac{\prod_{i=1}^{L^\prime} \omega_{\cal C}^{(i)}} {\prod_{i=1}^{L^\prime-1} 
\omega_\perp^{(i)}}  \exp  \left\{ -\frac{W_{\cal C}}{k_B T} \right\}
\quad .
\label{escape_rate}
\end{equation}
This formula, proposed by Langer \cite{langer,review}, is obtained by
considering the harmonic approximation of the potential in the vicinity of
${\cal C}_1$, ${\cal C}_2$, and ${\cal C}_s$. The $\omega_{\cal C}^{(i)}$'s are the
$L^\prime=2L-3$ non zero frequencies of the minimum $\cal C$
($\omega_{\cal C}^{(i)} = \sqrt{-\Lambda_{\cal C}^{(i)}}$, where
$\Lambda_{\cal C}^{(i)}$ is the $i$-th negative eigenvalue of the Hessian of the
potential energy $V$). Analogously, $\omega_\perp^{(i)}$'s are the $L^\prime-1$
non zero frequencies of the saddle (those corresponding 
to the stable directions), while $\omega_\parallel$ is associated with
the only expanding direction. Finally, $\gamma$ is the dissipation rate
\cite{nota1}, while the Arrhenius exponential factor depends on the height of 
the barrier, $W_{\cal C} = V_s - V_{1,2}$, normalized to the reduced temperature
$K_B T$, $K_B$ being the Boltzmann constant.

The above expression has been shown to reproduce reasonably well the numerically
obtained escape rates for heteropolymers in 2d at moderate temperatures for
$T < \approx T_\theta$ \cite{tap}. Accordingly, in that regime, both the folding
and the unfolding dynamics towards and from the NC are driven by thermal
activation processes, which determine the transitions between DCP. Small
$\Gamma$-values suggest that the heteropolymer may be trapped into some local
valley of the EL, far from the NC. Eq.~(\ref{escape_rate}) shows explicitly
that in high--dimensional spaces, the escape rate does not simply depend on the
energy barriers but also on entropic factors, which depend on geometrical
features of basins of attractions of the stationary configurations in the EL.

\begin{figure}[ht!]
\includegraphics[clip,width=6.5cm]{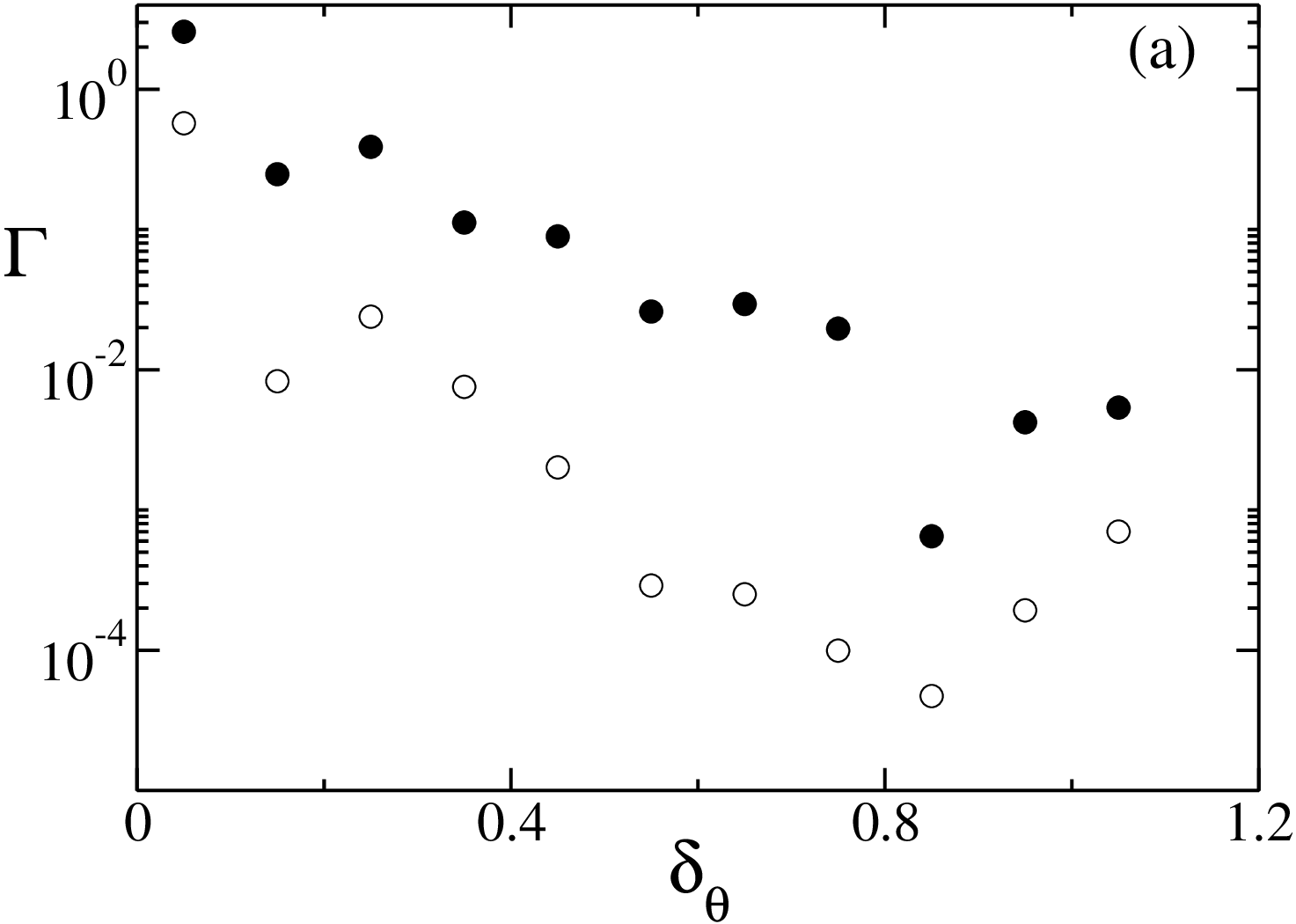}
\includegraphics[clip,width=6.5cm]{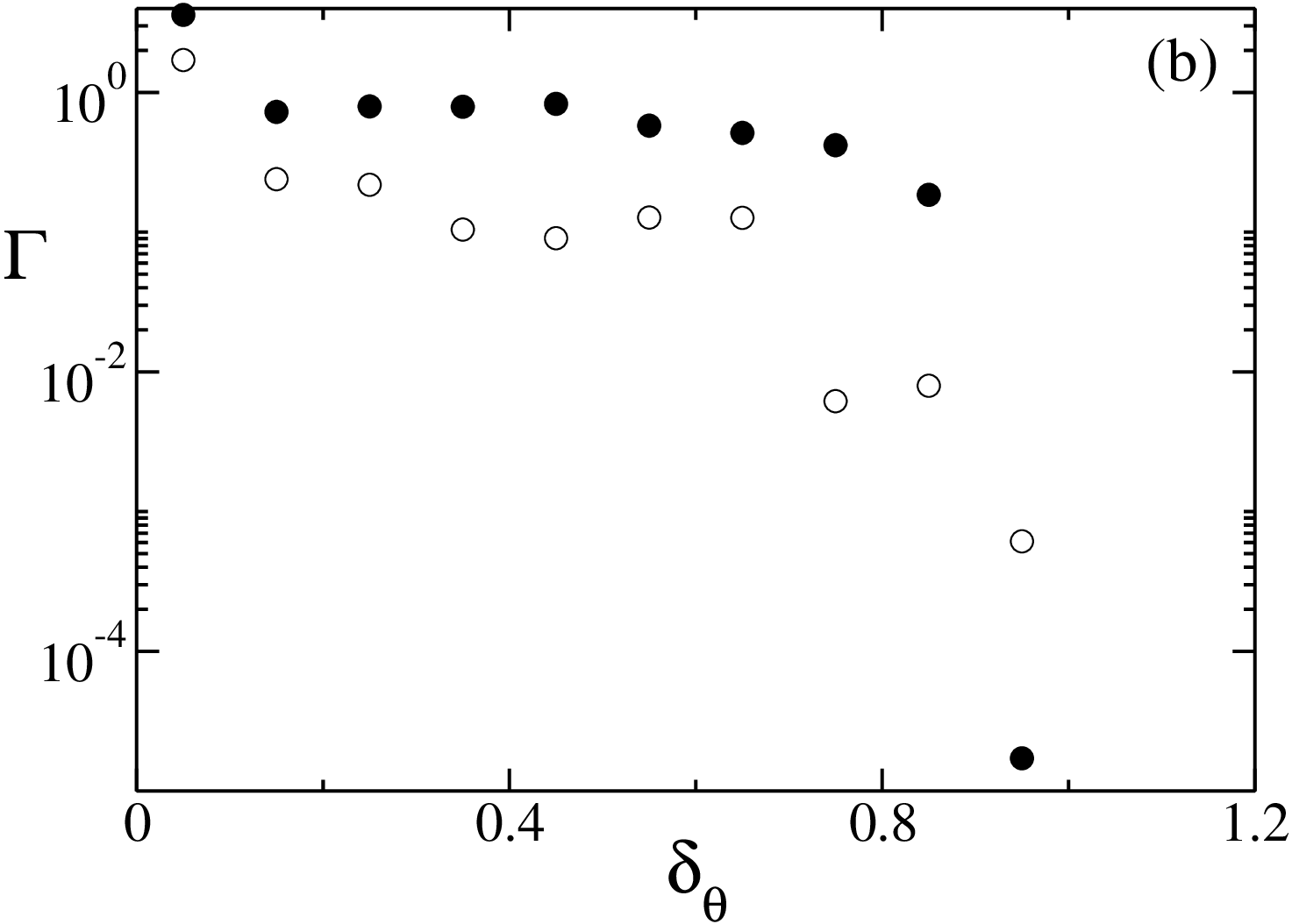}
\includegraphics[clip,width=6.5cm]{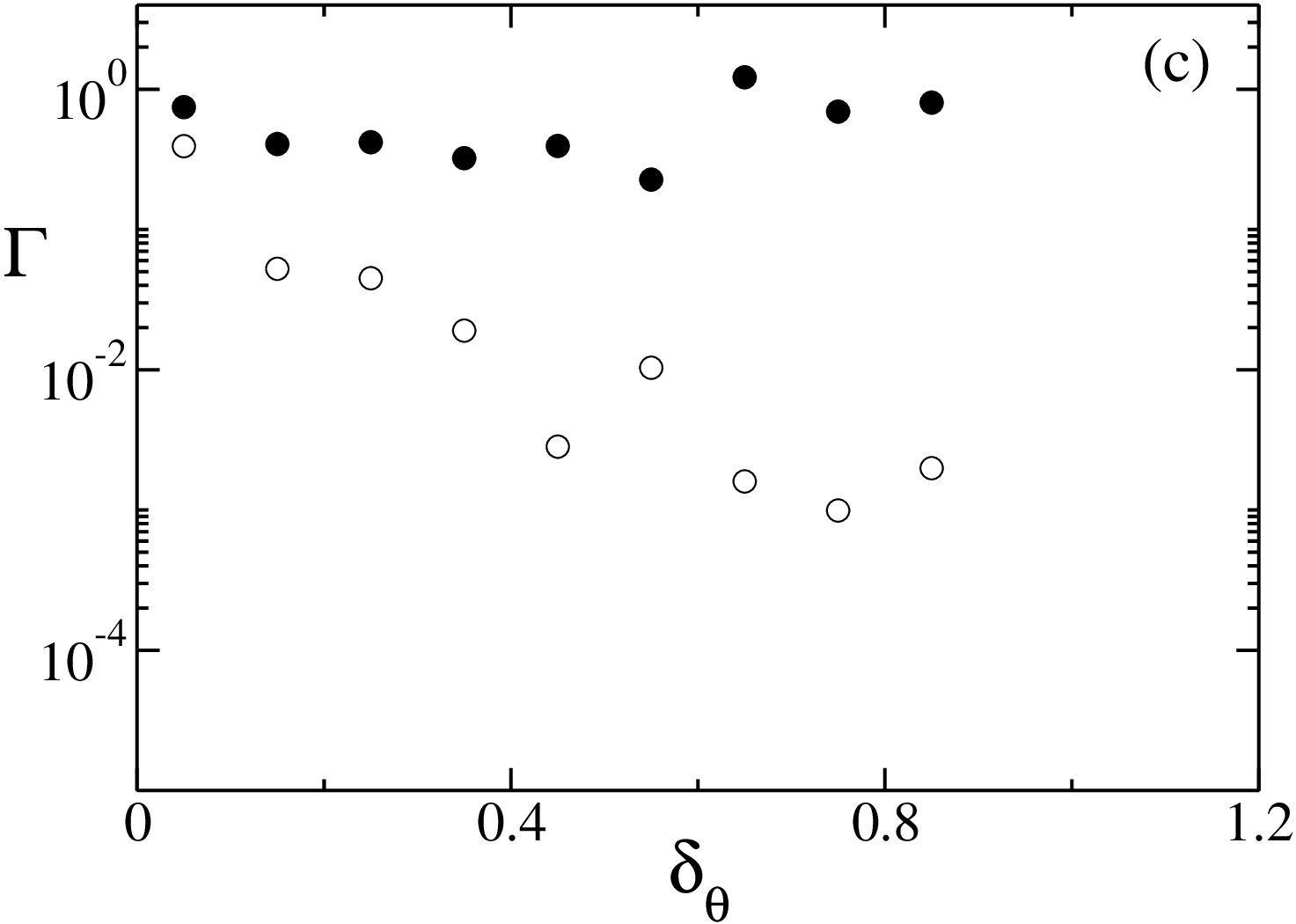}
\caption{Escape rates (\protect\ref{escape_rate}) 
versus the angular distance 
$\delta_\Theta$ for each DCP for sequence S0 (a) and S1 (b) and S2 
(c) at a temperature $T=T_f$ (namely, $T_f=0.044$ for S0 and S2, $T_f=0.061$ 
for S1). The filled circles refer to the transition rates 
$\Gamma_{{\cal C}_1}$, while the empty circles to $\Gamma_{{\cal C}_2}$.
}
\label{fig:w_entro}
\end{figure}

In Fig.~\ref{fig:w_entro} we have plotted the rates $\Gamma_{\cal C}$ as a
function of $\delta_\theta$ for the three sequences at $T=T_f$ and $\gamma =1$
(since past simulations indicate that the effective value of $\gamma$ is the
same at least in the whole NV, there is no need to know it when a comparative
analysis is being carried out). There we notice a striking difference between
S0 and S1 at large distances: actually the escape rates of S0 are two orders of
magnitude smaller than those of S1. Moreover, $\Gamma_{{\cal C}_2} \sim {\cal O}(1)$
almost in the whole range of distances $\delta_\theta$ for S1, indicating that no
trapping is expected in the shallower minima, while it exhibits an almost
exponential decrease for S0. An intermediate situation is observed for S2
at large distances. As a result, we see that a true funnel-like structure is
markedly present only in the EL of the sequence S1, that was indeed already
identified as a good folder by looking at different indicators
\cite{irback1,tlp}.

\section{Conclusions and perspectives}
\label{cinque}

In this paper we have analyzed the structure of the energy landscape (EL)
of a 2d off-lattice model of a polypeptidic chain and found that the relative
closeness between neighbouring minima can be exploited to implement an
effective algorithm to identify directly connected minima. In order to put the
analysis on firm quantitative grounds, we have tested several definitions of 
distance between configurations, finding that the best performances are obtained
for the angular distance $\delta_\theta^{(1)}$ (see Eq.~(\ref{distang})).
In fact, the $\delta_\theta^{(1)}$ distances corresponding to DCP are more sharply
concentrated at small values than for all other definitions of distances. In
particular, we have found that restricting the systematic search to all pairs
of minima whose distance is smaller than $\delta_\theta^{(1) \star} = 0.5$, one
can recover almost $99\%$ of all DCP.
The drawback of this approach is that a tiny fraction of DCP is unavoidably
missed, but the consequences are not a priori clear. Since our analysis of the
native valley has shown that minima are more rarefied in the vicinity of the
native configuration (see Fig.~\ref{fig:dist_nat}), we conclude that it is wise
to complement the above metric criterion with an extensive search limited to
the minima of the native valley. It is now important to verify to what extent
such a scenario extends to more realistic 3D systems, where the implementation
of effective algorithms to reconstruct the EL is even more crucial than in 2D.
Furthermore, important hints about the true relevance of missing links in a
connectivity graph will come after imposing a dynamics on the graph itself by
adding the activation rates $\Gamma$  relative to the transitions between
directly connected minima. By comparing the resulting evolution with that of
the original system one can in particular determine the minimal fraction of DCP
that is necessary to identify for a meaningful reconstruction of the folding
process.

Finally, in order to test how the observed rarefied density of minima in the
vicinity of the NC is connected to the presence of a true funnel-like structure,
we have computed the activation rate $\Gamma$  inside the NV. It turns out that
moving away from the NC, while in the homopolymer $\Gamma$ decreases very
rapidly, it stays almost constant in the sequence S1, previously identified
as a good foolder by other means. This is a clear indication that the
homopolymer can be trapped in several minima far from the minimal energy state,
while an accessible native valley does exist for S1.

\acknowledgments

We acknowledge CINECA in Bologna and INFM for providing us access to the
Beowulf Linux-cluster under the grant ,,Iniziativa Calcolo Parallelo''.
This work has been partially supported by a grant of the Ente Cassa di
Risparmio di Firenze, Italy, by the European Community via the STREP project
EMBIO (NEST contract N. 12835) and under the Italian FIRB project RBAU01BZJX 
``Dynamical and statistical analysis of biological microsystems''.
We also want to thank Dr. M. Riccardi for useful discussions and suggestions.

\vskip 0.5 truecm
\appendix{\bf Appendix}
\vskip 0.5 truecm

The database containing the minima of a model sequence can be used to determine
the first--order saddles directly connecting neighboring minima. The method we
propose to accomplish this goal is based on two different algorithms to solve
the following two problems: 
\begin{itemize}
\item Given two minima ${\cal C}_1$ and ${\cal C}_2$ and two configurations
${\cal P}_1$ and ${\cal P}_2$ belonging to their basins of attraction, one wants to
determine two configurations ${\cal Q}_1$ and ${\cal Q}_2$ on the segment
${\cal P}_1 {\cal P}_2$ arbitrarily close to the ridge dividing the two basins and
lying on its opposite sides; 

\item given ${\cal Q}_1$ and ${\cal Q}_2$, as defined above, one wants to apply an
iterative procedure, leading the two points to converge on the saddle.
\end{itemize}

The procedure is here described in more details:

\begin{enumerate}
\item given ${\cal Q}_1 = {\cal P}_1$ and ${\cal Q}_2= {\cal P}_2$, an intermediate
configuration ${\cal C}$ is defined, by setting its bond angles equal to the
average of the corresponding angles of ${\cal Q}_1$ and ${\cal Q}_2$;

\item a steepest descent procedure is applied to ${\cal C}$ until a minimum
${\cal C}_m$ is reached;

\item 
\begin{enumerate}
\item if ${\cal C}_m$ coincides with ${\cal C}_1$ (${\cal C}_2$), we replace
${\cal Q}_1$ (${\cal Q}_2$) with ${\cal C}$ and repeat the step (2) until the
euclidean distance $d_E({\cal Q}_1,{\cal Q}_2)$ between ${\cal Q}_1$ and
${\cal Q}_2$ becomes smaller than a given threshold $\delta$;

\item if ${\cal C}_m$ differs from both ${\cal C}_1$ and ${\cal C}_2$, we assume that
the basins of attraction of the two minima are not directly connected (i.e.,
${\cal C}_1$ and ${\cal C}_2$ are not neighbors) and ${\cal C}_m$ is added to the
minima database, provided it is not already included;

\item if $d_E({\cal Q}_1,{\cal Q}_2)< \delta$, we conclude they are close enough to
the ridge (i.e. the stable manifold separating the basins of attraction of
${\cal C}_1$ and ${\cal C}_2$) and pass to item (4);
\end{enumerate}

\item we let ${\cal Q}_1$ and ${\cal Q}_2$ evolve in time according to a steepest
descent relaxation until $d_E({\cal Q}_1(t),{\cal Q}_2(t))$ becomes larger than
$\delta$ and go back to (1), identifying ${\cal P}_1$ with ${\cal Q}_1(t)$ and
${\cal P}_2$ with ${\cal Q}_2(t)$;

\item when, during the steepest descent, the gradient of the potential both in
${\cal Q}_1$ and ${\cal Q}_2$  becomes smaller than another given threshold, we
assume ${\cal Q}_1$ and ${\cal Q}_2$ to be close enough to a first order saddle and
refine the configuration by means of a Newton's algorithm.
\end{enumerate}

It must be stressed that this algorithm not only allows to find first order
saddles but also enriches the number of known minima. The saddle-searching
strategy here proposed can then be viewed as an ``all purpose" method suitable
for a complete exploration of all the features of the EL relevant for protein
dynamics. 

The data reported in this paper were obtained by applying the algorithm to an 
initial database of minima determined by the method described in
Ref.~\cite{tap}.  It amounts to performing a high-temperature (typically above
$T_f$) Langevin dynamics, when the system is expected to visit a large portion
of the accessible phase space. The resulting trajectory is then sampled to
pinpoint a series of configurations, which are afterwards relaxed according to
a steepest descent dynamics and finally refined by means of a standard Newton's
method. This procedure was used for identifying a small database of 50-100
collapsed states for each  of the three sequences described in section
\ref{due}.

All possible pairs in these initial minima databases were then searched for DCP
by means of our saddle-finding algorithm. The new minima found during each run
of the algorithm on all possible pairs were stored in the database to
be investigated in successive runs. Actually the number of pairs of minima
grows much faster than the number of pairs investigated, thus making
impossible a complete analysis. The number of minima and saddles that were
identified after three runs is reported in Table~\ref{statistica}. 

In order to perform a complete search of all DCP at least in a restricted set
of minima, we have selected all configurations of energy lower
than  $V_f=V_0 + L T_f$. In this restricted database all pairs of minima
characterized by an angular distance smaller than $0.5$ where
investigated. The total number of saddles found by this 
procedure is reported in Table~\ref{statistica}.

\begin{table}[ht]
\caption[tabone]{
Number of minima in the whole database and in the first and second shell.
Number of investigated pairs and number of DCP found. 
Data have been reported for all the three examined sequences.
}
\begin{ruledtabular}
\begin{tabular}{lrrrrr}
 \hfil & \hfil S0 \hfil & \hfil S1 \hfil &
\hfil S2 \hfil \hfil \\
\hline
\begin{tabular}{l} Total number of \\ minima \end{tabular} &  \,\,\,\,
     $2.3 \times 10^5$ &  \,\,\,\, $7.2 \times 10^4$ & \,\,\,\,$6.4 \times 10^4$ \\
\hline
\begin{tabular}{l} Number of minima \\ in the 1st shell \end{tabular} &  \,\,\,\,
     64  &  \,\,\,\,50 &  \,\,\,\,49 \\
\hline
\begin{tabular}{l} Number of minima \\ in the 2nd shell \end{tabular} &  \,\,\,\,
     1,181  &  \,\,\,\,465 &  \,\,\,\,348 \\
\hline
\begin{tabular}{l} Number of \\ minima below $T_f$ \end{tabular} &  \,\,\,\,
     66,470 &  \,\,\,\,5,883 & \,\,\,\,8,670 \\
\hline
\begin{tabular}{l}Number of \\
investigated pairs\end{tabular} &$9.1\times 10^6$& $0.94 \times 10^6$ &
        $1.3 \times 10^6$\\
\hline
\begin{tabular}{l}Number of \\ connected pairs\end{tabular} & 84,990 
& 10,470 & 14,356\\
\end{tabular}
\end{ruledtabular}
\label{statistica}
\end{table}



\begin{thebibliography}{9}
\bibitem{nota1} This is a free parameter, which has to be fixed 
according to some physical criterion.
\end{thebibliography}
\end{document}